\documentstyle[aps,preprint,prl]{revtex}
\begin{document}

\title{A Hartree--Bose Mean-Field Approximation for the Interacting Boson
Model (IBM-3)}
\author{
J.E.~Garc\'{\i}a--Ramos$^1$,
J.M.~Arias$^1$,
J.~Dukelsky$^2$,
E.~Moya de Guerra$^2$
and P.~Van Isacker$^3$ }
\address{$^1$Departamento de F\'{\i}sica At\'omica, Molecular y Nuclear,
Universidad de Sevilla, Aptdo. 1065, 41080 Sevilla, Spain}
\address{$^2$Instituto de Estructura de la Materia,
Serrano 123, 28006 Madrid, Spain}
\address{$^3$Grand Acc\'el\'erateur National d'Ions Lourds,
B.P.~5027, F-14076 Caen Cedex 5, France}
\date{\today}
\maketitle

\begin{abstract}
A Hartree--Bose mean-field approximation for the IBM-3 is presented.
A Hartree--Bose transformation from spherical to deformed bosons
with charge-dependent parameters is proposed
which allows bosonic pair correlations
and includes higher angular momentum bosons.
The formalism contains
previously proposed IBM-2 and IBM-3 intrinsic states
as particular limits.
\end{abstract}


\draft
With the advent of radioactive nuclear beam (RNB) facilities
unexplored regions of the nuclear chart
will become available for spectroscopic studies.
New aspects of nuclear dynamics
and novel types of collectivity and nuclear topologies
are expected.
Nuclei with roughly equal numbers of protons and neutrons ($Z\sim N$)
and with masses in between $^{40}$Ca and $^{100}$Sn
are of particular interest
because the build-up of nuclear collectivity in this mass region
occurs in the presence of pairing correlations
between alike nucleons (proton--proton and neutron--neutron)
{\em as well as} between neutrons and protons \cite{Warn94}.
This offers the possibility of experimentally accessing nuclei
that exhibit a superconducting phase
arising from proton--neutron Cooper pairs \cite{Naza97}.
Although recent breakthroughs \cite{Koon97,Honm96}
have made shell-model calculations possible for this mass region,
they are still of a daunting complexity
and alternative approximation schemes are required
that yield a better intuitive (e.g.\ geometric) insight.

One of the possible alternatives
is the Interacting Boson Model (IBM) \cite{Iach87}.
It has been shown \cite{El80}
that nuclei with protons and neutrons filling the same valence shell
require an extended boson model, IBM-3.
In IBM-3 three types of bosons are included:
proton--proton ($\pi$),
neutron--neutron ($\nu$),
and proton--neutron ($\delta$).
The $\pi$, $\nu$, and $\delta$ bosons
are the three members of a $T=1$ triplet,
and their inclusion is necessary to obtain
an isospin-invariant formulation of the IBM.
Over the last decade the validity of the IBM-3 has been tested
and its relationship with the shell model worked out
\cite{Ell87}-\cite{Lac95}.

The mean-field formalism has been an important tool
to acquire a geometric understanding of the IBM ground state
and of the vibrations around the deformed equilibrium shape
\cite{gino80}-\cite{Duke84}.
Moreover, a treatment based on mean-field techniques
generally leads to a considerable reduction
in the complexity of the calculation,
allowing the introduction of additional degrees of freedom if needed.
Studies in the intrinsic framework are thus useful to assess
the importance of higher angular momentum bosons
with e.g.\ $\ell=3^-,4^+,\dots$
or to investigate the role of extra degrees of freedom not included in IBM
such as two-quasiparticle excitations, etc.

An intrinsic-state formalism for the IBM-3
was recently presented by Ginocchio and Leviatan (GL) \cite{gino94}.
In that work charge-independent deformation parameters
are imposed in the Hartree--Bose transformation
from spherical to axially deformed bosons
and the trial wavefunction is taken to have good isospin.
Closer inspection reveals that this trial wavefunction
has the additional isospin SU(3) symmetry
which in IBM-3 is equivalent to orbital U(6) symmetry.
(Isospin SU(3) symmetry is to IBM-3
what $F$-spin symmetry \cite{Otsu78} is to IBM-2.)
Moreover, it is well known that isospin symmetry itself
is increasingly broken in $Z\sim N$ nuclei
as the nuclear mass increases \cite{Colo95}.
There is also tentative evidence for rigid triaxial shapes in the 
region of interest and its proper description would require the inclusion 
of three body forces or higher angular momentum bosons.
We therefore present in this paper
a generalization of the treatment of GL
in which none of the above symmetries
(isospin SU(3) and SU(2))
is imposed on the trial wavefunction
and which includes bosons of angular momenta higher than $\ell=2$.
For practical applications we restrict ourselves here to $\ell=0,2$.

We start with the usual spherical boson creation and annihilation operators
$\gamma _{\ell m\tau}^{\dagger},\;\gamma _{\ell m\tau}$
where $\ell$ is the angular momentum,
$m$ is its third component,
and $\tau$ is the isospin projection.
Each boson carries isospin $T=1$.
We also define
$\tilde{\gamma}_{\ell m\tau}=(-1)^{\ell-m}\gamma _{\ell-m\tau}$.
In terms of these boson operators,
a system of $N$ bosons interacting
through a general number-conserving two-body Hamiltonian
can be written in multipolar form as
\begin{equation}
\label{H1}
H=
\sum\limits_{\ell \tau}
\varepsilon_{\ell \tau}
\gamma^\dagger_{\ell\tau}\cdot\tilde{\gamma}_{\ell\tau}+
\sum_{L}\sum\limits_{\tau_1\tau_2\tau_3\tau_4}
\kappa^L_{\tau_1\tau_2\tau_3\tau_4}
\hat{T}^L_{\tau_1\tau_2}\cdot\hat{T}^L_{\tau_3\tau_4},
\end{equation}
where the symbol $\cdot$ denotes scalar product in orbital space.
In isospin space the only restriction is
$\tau_1+\tau_2=\tau_3+\tau_4$
(i.e., a charge-conserving Hamiltonian is assumed)
and $\hat{T}^L_{\tau_1\tau _2}$ are multipole operators
with total angular momentum $L$,
\begin{equation}
\label{OpeMul}
\hat{T}^L_{M,\tau _1\tau _2}=
\sum_{\ell_1\ell_2}\;
\chi_{\ell_1\ell_2,\tau_1\tau_2}^L
(\gamma^\dagger_{\ell_1\tau_1}\times
\tilde{\gamma}_{\ell_2\tau_2})^L_M,
\end{equation}
where the coupling is only done in angular momentum.
The Hamiltonian (\ref{H1}) can be used for IBM-3, for IBM-2,
or even for a general isospin non-conserving Hamiltonian
with three kinds of bosons.

Deformed bosons are defined in terms of spherical ones
by means of a unitary Hartree--Bose transformation
\begin{equation}
\label{Har1}
\Gamma _{p\tau }^\dagger=
\sum_{\ell m}\eta_{\ell m}^{p\tau}\gamma_{\ell m\tau}^\dagger,
\qquad
\gamma_{\ell m\tau}^\dagger=
\sum_p\eta_{\ell m}^{*p\tau}\Gamma_{p\tau}^\dagger,
\end{equation}
and their hermitian conjugates.
The deformation parameters $\eta_{\ell m}^{p\tau}$
in these equations verify the orthonormalization conditions
\begin{equation}
\label{orto}
\sum_{\ell m}
\eta_{\ell m}^{*p'\tau}
\eta_{\ell m}^{p\tau}
=\delta_{pp'},
\qquad
\sum_{p}
\eta_{\ell m}^{*p\tau }
\eta_{\ell'm'}^{p\tau}
=\delta_{\ell\ell'}
\delta_{mm'}.
\end{equation}
Note the explicit dependence
on the isospin component $\tau$ of the transformation $\eta$,
allowing different structures for the different condensed bosons
$\pi$, $\nu$, and $\delta$.
The index $p$ labels different possible deformed bosons.
We choose $p=0$ for the fundamental deformed bosons
and $p=1,2,\dots$ for the different excited bosons.
For instance, in an SU(3) scheme different values of $p=0,1,2,3$
label the ground, $\beta$, $\gamma$, and scissors bands, respectively.
Since in this work we only treat the ground-state condensed boson,
the Hartree superscript $p$ is always zero here and it will be 
omitted in the following.
The formalism for the excited states will be presented elsewhere.

Following Ref.~\cite{gino94},
the trial wavefunction for the ground state of an even--even system
with a proton excess is of the form
(the trial wavefunction for an even--even system
with a neutron excess is obtained
by interchanging the role of protons and neutrons)
\begin{equation}
\label{Gswf}
\left|\phi(\alpha)\right\rangle
={\Lambda^\dagger}^{N_n}(\alpha)
{\Gamma_{1}^\dagger}^{N_p-N_n}
\left|0\right\rangle,
\end{equation}
where the operator $\Lambda^\dagger$
creates a correlated bosonic pair in isospin space
\begin{equation}
\label{Pair}
\Lambda^\dagger(\alpha)
=\Gamma_{1}^\dagger\Gamma_{-1}^\dagger
+\alpha\Gamma_{0}^\dagger\Gamma_{0}^\dagger.
\end{equation}
In Eq.~(\ref{Gswf}) $N_p$ ($N_n$)
is the number of proton (neutron) pairs in the valence space.
The trial wavefunction (\ref{Gswf}) 
contains the isospin-conserving formalism of GL and the IBM-2
as natural limits.
Two different values of $\alpha$ are connected with these limits.
For $\alpha=-\frac 1 2$, $\Lambda^\dagger(\alpha)$ corresponds,
in the particular case of $\tau$-independent deformation parameters,
to an isoscalar bosonic pair.
Its total isospin is $T=N_p-N_n$
and the results of GL are reproduced.
Any other value of $\alpha$ breaks isospin symmetry.
In particular, $\alpha=0$ eliminates the mixing of $\delta$ bosons
in the ground state and yields an IBM-2 intrinsic state. 
It should be emphasized that when the deformation parameters 
$\eta_{\ell m}^{p\tau}$ in Eq. (3) depend on the isospin component $\tau$, 
the set of operators $\Gamma_{p \tau}^\dagger$ with $\tau=-1,0,1$ do not 
form an isospin triplet, and consequently the $\Lambda^\dagger$ in Eq. (6) 
may contain mixtures of T=0,1,2 isospin components.

At this point we would like to remark that the trial wavefunction 
(\ref{Gswf}) is not the most general U(18) intrinsic state, involving a 
combination of all $\tau=-1,0,1$ condense bosons. The U(18) intrinsic state 
is written as 
\begin{equation}
\label{U18}
\left|\phi\right\rangle_{U(18)}
=({\Gamma_{c}^\dagger})^{N_p+N_n}
\left|0\right\rangle,
\end{equation}
where
\begin{equation}
\label{U18b}
\Gamma_c^\dagger =\sum_{\ell m \tau} \xi_{\ell m \tau}
\gamma_{\ell m \tau}^\dagger.
\end{equation}
In this state orbital angular momemtum, isospin, and charge are broken. The 
state given in Eq. (\ref{Gswf}) improves over this state by including
charge conserving pair correlations. Thus, it is expected to lead to a
deeper energy minimum. Numerical results illustrating this point will be 
presented later on.

The variational parameters of the trial wavefunction
are the matrix elements $\eta_{\ell m}^{\tau}$
of the Hartree--Bose transformation,
associated with the orbital and isospin degrees of freedom,
and the parameter $\alpha$,
which determines the amount of mixing of $\delta$ bosons
in the ground state.

The ground-state energy is obtained
by taking the expectation value of the Hamiltonian (\ref{H1})
in the state (\ref{Gswf}):
\begin{equation}
\label{Ener}
E(\eta,\alpha)=
\sum_\tau\epsilon_\tau f_1(\alpha,\tau)+
\sum_{\tau_1\tau_2\tau_3\tau_4}
V_{\tau_1,\tau_2,\tau_3,\tau_4}^c
f_2(\alpha,\tau_1\tau_2\tau_3\tau_4),
\end{equation}
where
\begin{equation}
\label{Htauinf1}
\epsilon_\tau=
\sum_{\ell m}\tilde \varepsilon_{\ell\tau}
\eta_{\ell m}^{*\tau}
\eta_{\ell m}^{\tau},
\end{equation}
\begin{equation}
\label{Htauinf2}
\begin{array}{c}
V_{\tau_1,\tau_2,\tau_3,\tau_4}^c=
\sum\limits_{\ell_1m_1\ell_2m_2\ell_3m_3\ell_4m_4}
{\displaystyle
{V_{\ell_1m_1\tau_1,\ell_2m_2\tau_2,\ell_3m_3\tau_3,\ell_4m_4\tau_4}}}
~ \eta_{\ell_1m_1}^{*\tau_1}
\eta_{\ell_2m_2}^{*\tau_2}
\eta_{\ell_3m_3}^{\tau_3}
\eta _{\ell_4m_4}^{\tau _4},
\\
\end{array}
\end{equation}
\begin{equation}
\label{f1}
f_1(\alpha,\tau)=
\frac{\left\langle\phi(\alpha)
\right|\Gamma_{\tau}^\dagger\Gamma_{\tau}\left|
\phi(\alpha)\right\rangle}
{\langle\phi(\alpha)\mid\phi(\alpha)\rangle},
\end{equation}
and
\begin{equation}
\label{f2}
f_2(\alpha,\tau_1\tau_2\tau_3\tau_4)=
\frac{\left\langle\phi(\alpha)
\right|\Gamma_{\tau_1}^\dagger\Gamma_{\tau_2}^\dagger
\Gamma_{\tau_3}\Gamma_{\tau _4}\left|
\phi(\alpha)\right\rangle}
{\langle\phi(\alpha)\mid\phi(\alpha)\rangle}.
\end{equation}
The coefficients $\tilde \varepsilon_{\ell\tau}$ include the single particle 
energies $\varepsilon_{\ell\tau}$ in Eq. (\ref{H1}) plus  
contributions from the two body term in the same equation.
The coefficients
$V_{\ell_1m_1\tau_1,\ell_2m_2\tau_2,\ell_3m_3\tau _3,\ell_4m_4\tau _4}$
are the symmtrized interaction matrix elements between normalized two-boson 
states following Ref. \cite{Duke84},
\begin{equation}
\begin{array}{l}
V_{\ell_1m_1\tau_1,\ell_2m_2\tau_2,\ell_3m_3\tau_3,\ell_4m_4\tau_4}\equiv
{1 \over 4}~ \left\langle\ell_1m_1\tau_1,\ell_2m_2\tau_2\right|H
\left|\ell_3m_3\tau_3,\ell_4m_4\tau_4\right\rangle~ \\ \\
~~~~~~~~~~~~~~~~~~~~~~~~~~~~~~~~~~~~~~\times ~
\sqrt{1+\delta_{\ell_1\ell_2}\delta_{m_1 m_2}\delta_{\tau_1\tau_2}}
\sqrt{1+\delta_{\ell_3\ell_4}\delta_{m_3 m_4}\delta_{\tau_3\tau_4}}
\end{array}
\end{equation}

The dependence of the energy on the variational parameters $\eta $'s
is contained in the one-body $\epsilon$ (\ref{Htauinf1})
and the two-body $V^c$ (\ref{Htauinf2}) terms,
while the dependence on $\alpha$ comes through
the isospin matrix elements $f_1$ (\ref{f1}) and $f_2$ (\ref{f2}).
The latter matrix elements are straightforward to calculate
by a binomial expansion of the ground-state trial wavefunction (\ref{Gswf}).

The Hartree--Bose equations for the orbital variational parameters $\eta $
are obtained by minimizing the energy (\ref{Ener})
constrained by the norm of the transformation.
Assuming a charge-conserving Hamiltonian (\ref{H1})
the following Hartree--Bose equations result:
\begin{equation}
\label{Har2}
\sum_{\ell_2m_2}
h_{\ell_1m_1,\ell_2m_2}^\tau
\eta_{\ell_2m_2}^{\tau}=
E_\tau\eta_{\ell_1m_1}^{\tau},
\end{equation}
where the Hartree--Bose matrix $h^\tau$ is
\begin{equation}
\label{Har3}
\begin{array}{l}
h_{\ell_1m_1,\ell_2m_2}^\tau
=\epsilon_{\ell_1\tau}f_1(\alpha,\tau)
\delta_{\ell_1\ell_2}\delta_{m_1m_2}
\\\\
\qquad
+2\sum\limits_{\ell_3m_3\ell_4m_4\tau_2\tau_3\tau_4}
{\displaystyle
{V_{\ell_1m_1\tau,\ell_3m_3\tau_3,\ell_4m_4\tau_4,\ell_2m_2\tau_2}}}
{\eta_{\ell_3m_3}^{*\tau_3}\eta _{\ell_4m_4}^{\tau _4}
 \eta_{\ell_2m_2}^{\tau_2}
\over
{\eta_{\ell_2m_2}^{\tau}}}
f_2(\alpha,\tau\tau_3\tau_4\tau_2).\\
\end{array}
\end{equation}
The term ${\eta_{\ell_2m_2}^{\tau}}$ in the denominator
is a consequence of a mathematical trick
for obtaining a set of three coupled Hartree--Bose equations (\ref{Har2}).
These depend on the isospin matrices $f_1$ and $f_2$.
For each value of $\alpha$
the matrices $f_1$ and $f_2$ are calculated
and the Hartree--Bose equations (\ref{Har2}--\ref{Har3})
are solved self-consistently.
The procedure is iterated
until one finds the absolute minimum of the energy (\ref{Ener}). 
Once the problem is solved self-consistently,
the diagonalization of (\ref{Har2}) provides
the deformation parameters $\eta_{\ell m}^{\tau}$ for the ground state.

To test the present formalism
and to compare with the one by GL,
we used a simple Hamiltonian recently proposed by Ginocchio \cite{gino96},
\begin{equation}
\label{ginham}
H=-\kappa\,\sum_{T=0,1,2}\hat{P}^T:\hat{P}^T,
\end{equation}
where
\begin{equation}
\label{ginham2}
\hat{P}^T=
(s^\dagger{\tilde{\tilde d}}
+(-1)^T d^\dagger{\tilde{\tilde s}})^{L=2,T}.
\end{equation}
In these equations
the symbol $:$ denotes a scalar product in orbital and isospin spaces
and ${\tilde{\tilde\gamma}}_{\ell m\tau}=
(-1)^{\ell-m+1-\tau}\gamma_{\ell-m-\tau}$.
The Hamiltonian (\ref{ginham}) is clearly isospin invariant
and provides a first simple test to the present formalism

Figure~1 shows, for a system with 5 proton pairs and 3 neutron pairs,
the ground-state energy for the Hamiltonian (\ref{ginham})
as a function of $\alpha$.
The dashed line is calculated with $\tau$-independent deformation parameters;
the GL minimum energy is reproduced for $\alpha=-\frac 1 2$.
The full line is calculated with the present formalism.
The latter calculation always gives a lower energy and,
in particular, the minimum is not obtained for $\alpha=-\frac 1 2$
but for $\alpha\approx-0.32$.
In addition, the corresponding deformation parameters
are $\tau$ dependent. 
The energy gained by breaking isospin invariance in our trial 
wavefunction is relatively small. In this respect it may be advantageus to use
the GL intrisic state for isospin conserving hamiltonians. Though, a better 
approximation would be obtained by performing variation after isospin 
projection over our trial wavefunctions.

We note that for a system with equal number of protons and neutrons
the present formalism recovers exactly the GL results;
differences occur for $Z\ne N$.
This can be seen in Fig.~2
where the deformation parameters $\beta_\tau$
are plotted versus the difference $N_p-N_n$
(starting with 4 proton pairs and 4 neutron pairs).
The deformation parameters $\beta_\tau$ are obtained from
$\beta_\tau=\sqrt{{1\over\vert\eta^{0\tau}_{00}\vert^2}-1}$
(see Eq.~(1) of Ref.~\cite{gino94}).
For $N_p=N_n$ the deformation parameters are independent of $\tau$,
but not any longer as $N_p-N_n$ increases.
The proton and neutron deformations remain very close;
the $\delta$ deformation $\beta_\delta$, however,
quickly becomes very large in comparison.
This is because $N_{\delta}$ decreases as $N_p-N_n$ increases.
This effect can be seen in Fig.~3 where  
the mean values of the boson numbers, $N_\tau$, are plotted.
The same behaviour has been obtained recently with large scale shell
model calculations (see Ref.~\cite{engel96}).
In all our calculations we found that
the Ginocchio Hamiltonian (\ref{ginham})
leads to a $\gamma$-independent energy surface.   

It is worth noting that
the present formalism allows one to reproduce
the well-known case of triaxiality in IBM-2.
To show this we use the IBM-2 Hamiltonian
\begin{equation}
\label{hamibm2}
H= -(Q_{\pi}+Q_{\nu}^\prime)\cdot(Q_{\pi}+Q_{\nu}^\prime)\quad,
\end{equation}
where $\cdot$ denotes scalar product in angular momentum,
$Q_{\pi}$ is the SU(3) generator,
$Q=s^\dagger\tilde{d}+d^\dagger\tilde{s}-
\frac{\sqrt{7}}{2}(d^\dagger \times \tilde d)^{L=2}$,
for proton bosons
and $Q_{\nu}^\prime$ is the $\overline{\rm SU(3)}$ generator,
$Q^\prime=s^\dagger\tilde{d}+d^\dagger\tilde{s}+
\frac{\sqrt{7}}{2} (d^\dagger \times \tilde d)^{L=2}$,
for neutron bosons.
The minimization procedure now gives $\alpha=0$,
which corresponds to the IBM-2 limit.
In addition, the minimum deformation parameters
correspond to a prolate proton condensate,
axially symmetric about the intrinsic $z$ axis, 
and to an oblate neutron condensate
axially symmetric about the intrinsic $y$ axis,
giving rise to an overall triaxial shape. It should be pointed 
out that in this case there is no triaxial minimum for aligned 
proton--neutron shapes with equal deformations. Here the overall shape 
is triaxial but the underlying separate proton and neutron condensates
correspond to  different (prolate--oblate) axial shapes.

Finally, we present a calculation in which isospin is
explicitely broken by the hamiltonian: 

\begin{equation}
\label{hamnois}
H= \sum_{\ell \tau} \epsilon_{\ell \tau} \hat n_{\ell \tau}
       -{1 \over 5} ~ N[Q^0:Q^0 + {2 \over 3} Q^1:Q^1]
\end{equation}
where $Q^T=[s^\dagger{\tilde{\tilde{d}}}+d^\dagger{\tilde{\tilde{s}}}-
\frac{\sqrt{7}}{2}(d^\dagger \times {\tilde{\tilde d}})]^{L=2,T}$ and
$N[...]$ stands for normal ordering product.
In Figs. 4 and 5 we present the results of
a calculation with $\epsilon_{s \pi}=\epsilon_{s \nu}=0$,
$\epsilon_{d \pi}=\epsilon_{d \nu}=1.5 $,
$\epsilon_{s \delta}=2.3~ |N_p -N_n|$ and
$\epsilon_{d \delta}=1.5 + 2.3~ |N_p + N_n|$ (all $\epsilon$'s in MeV).
In Fig. 4 the deformation parameters, $\beta$'s, are shown as a funtion of
$N_p - N_n$ (the calculation starts with $N_p = N_n=4$ and then
$N_p$ is increased). Fig. 5 shows the corresponding ground state energies
for the intrinsic states of GL, and thos defined in eq. 5 and 7. It is 
interesting to note that the GL intrinsic state and our pair correlated 
intrinsic state produce the same results for $N_p - N_n = 0$, being better 
than the U(18) intrinsic state. For $N_p - N_n > 0 $  both isospin 
nonconserving intrinsic states give better results than GL. It is also 
interesting to see that our pair correlated ansatz is superior to the the 
U(18) for moderate values of $N_p - N_n$.      
No triaxial deformation is found in these calculations.

In summary, we have extended the intrinsic-state formalism
of Ginocchio and Leviatan \cite{gino94} for IBM-3
in three different ways.
First, the Hartree--Bose transformation is chosen to depend
on the isospin component $\tau$.
Second, variable isospin bosonic pair correlations are introduced
through the parameter $\alpha$.
Finally, higher-order bosons, other than the usual $s$ and $d$ bosons,
are included in the Hartree-Bose transformation.
This formalism contains the IBM-2 and GL intrinsic states
as particular limits.
Substantial differences in the deformation parameters
are obtained when $N_p \ne N_n$.  
We have presented results for isospin conserving and nonconserving 
hamiltonians with $s$ and $d$ bosons. Substantial differences in the 
deformation parameters $\beta_{\tau}$ are obtained for $N_p > N_n $.   
In most of the cases studied, our
pair correlated intrinsic ground states are lower in energy than the
GL ground states, although for the isospin conserving hamiltonian the 
energy gain is small. We therefore conclude that the new intrisic state
is useful for treating isospin breaking hamiltonians.

This work has been supported in part by the Spanish DGICYT under contracts
No. PB95/0123 and PB95--0533, a  DGICYT-IN2P3 agreement and by the 
European Commission under contract CI1*-CT94-0072.


\begin{figure}[]
\caption{Calculated ground-state intrinsic energy as a function of $\alpha$
for a system with 5 proton and 3 neutron pairs
interacting through the Ginocchio Hamiltonian  (\ref{ginham})
with $\kappa=1$ MeV.}
\end{figure}

\begin{figure}[]
\caption{Deformation parameters $\beta_\tau$
for a system with $N_n=4$ neutron pairs
as a function of the difference $N_p-N_n$
between the numbers of proton and neutron pairs.
The Ginocchio Hamiltonian  (\ref{ginham})  with $\kappa=1$ MeV is used.}
\end{figure}

\begin{figure}[]
\caption{Mean values of the boson numbers,  $N_\tau$,
for a system with $N_n=4$ neutron pairs
as a function of the difference $N_p-N_n$
between the numbers of proton and neutron pairs.
The Ginocchio Hamiltonian (\ref{ginham}) with $\kappa=1$ MeV is used.}
\end{figure}

\begin{figure}[]
\caption{Same as Fig. 2 but for the non-conserving isospin
Hamiltonian  (\ref{hamnois}) with the parameters given in the text.}
\end{figure}

\begin{figure}[] 
\caption{Calculated ground-state intrinsic energy, as a function of the 
difference $N_p-N_n$ between the numbers of proton and neutron pairs,
for the non-conserving isospin Hamiltonian  (\ref{hamnois}) with 
the parameters given in the text.}
\end{figure}

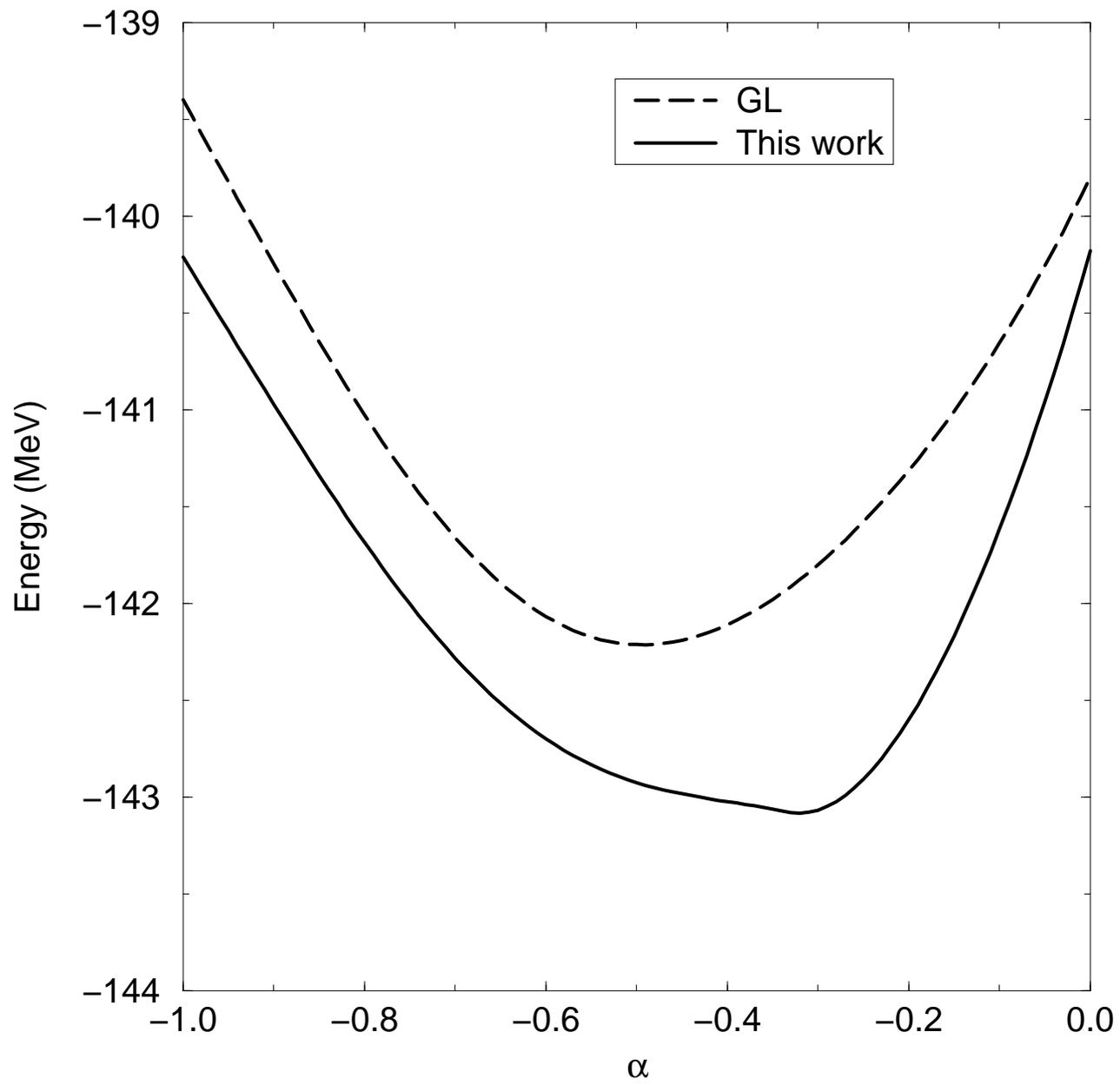
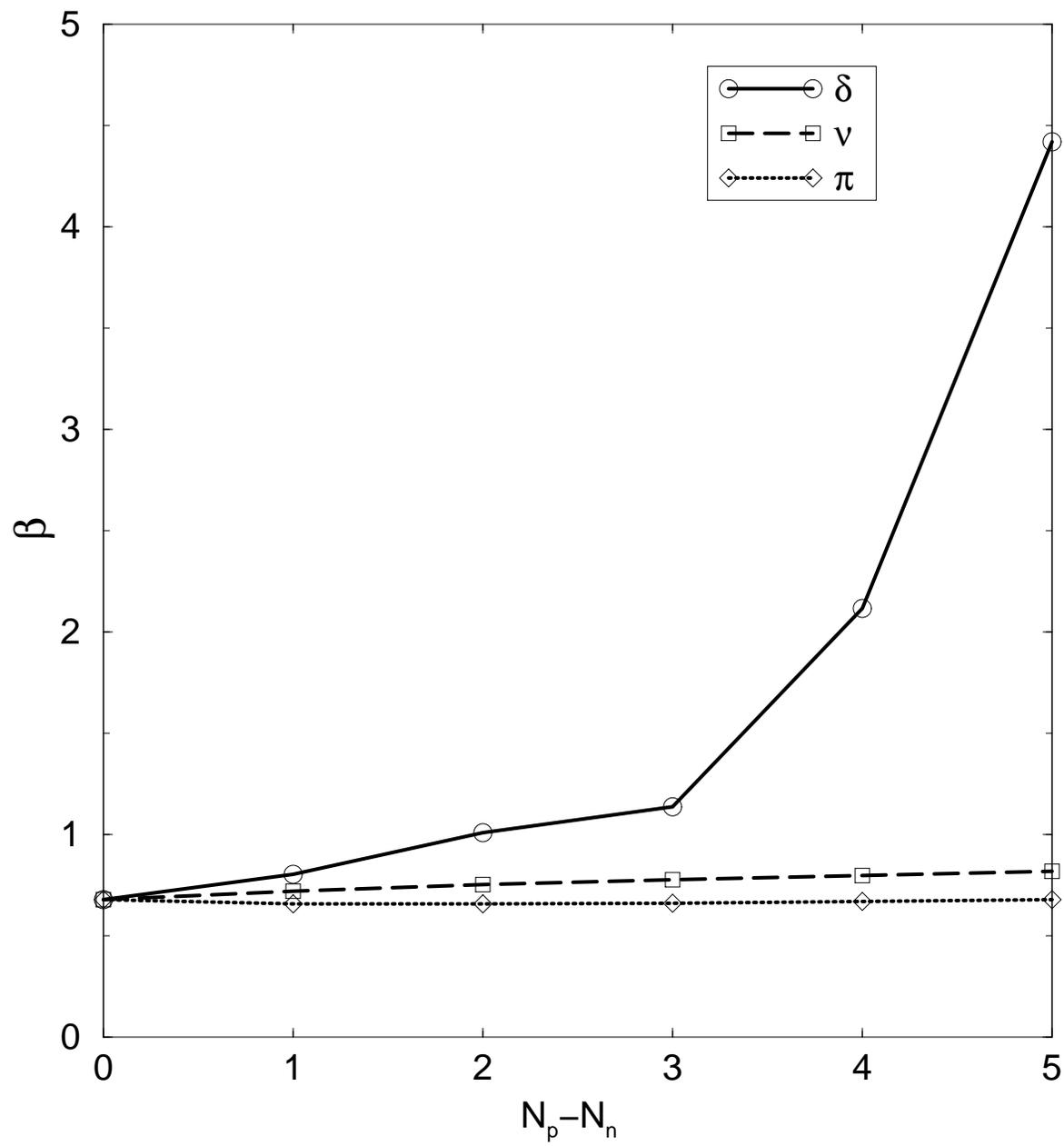
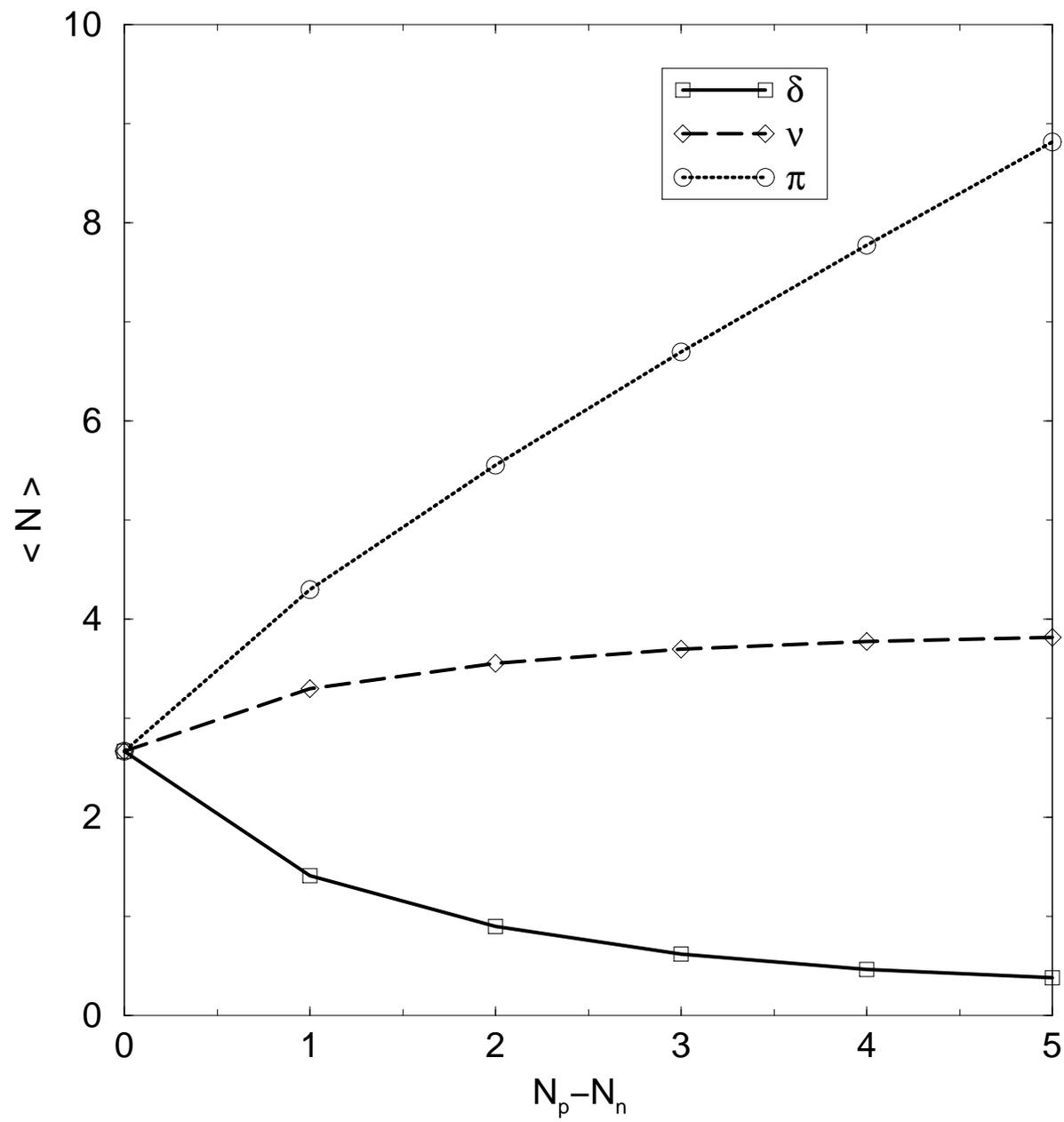
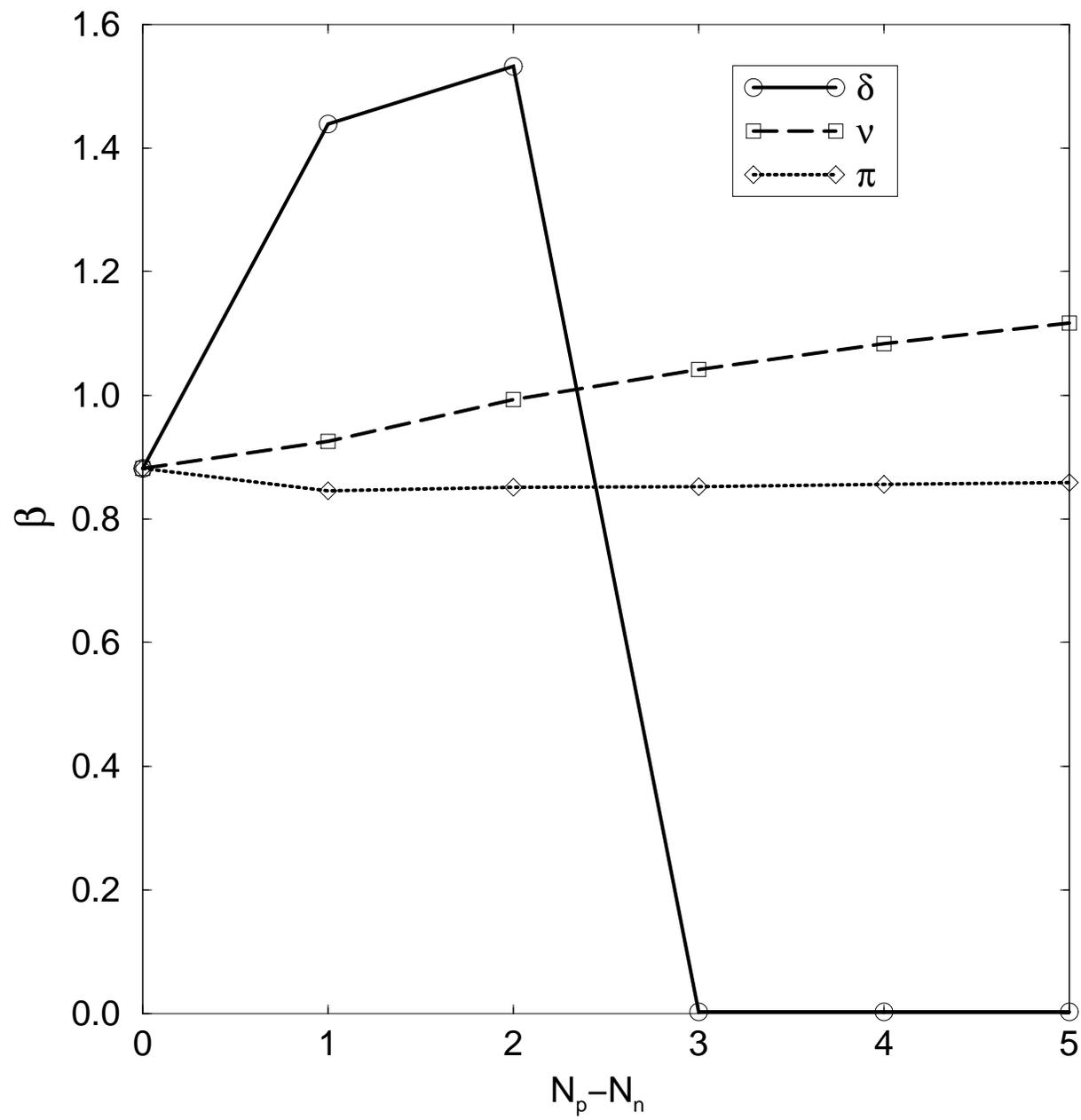
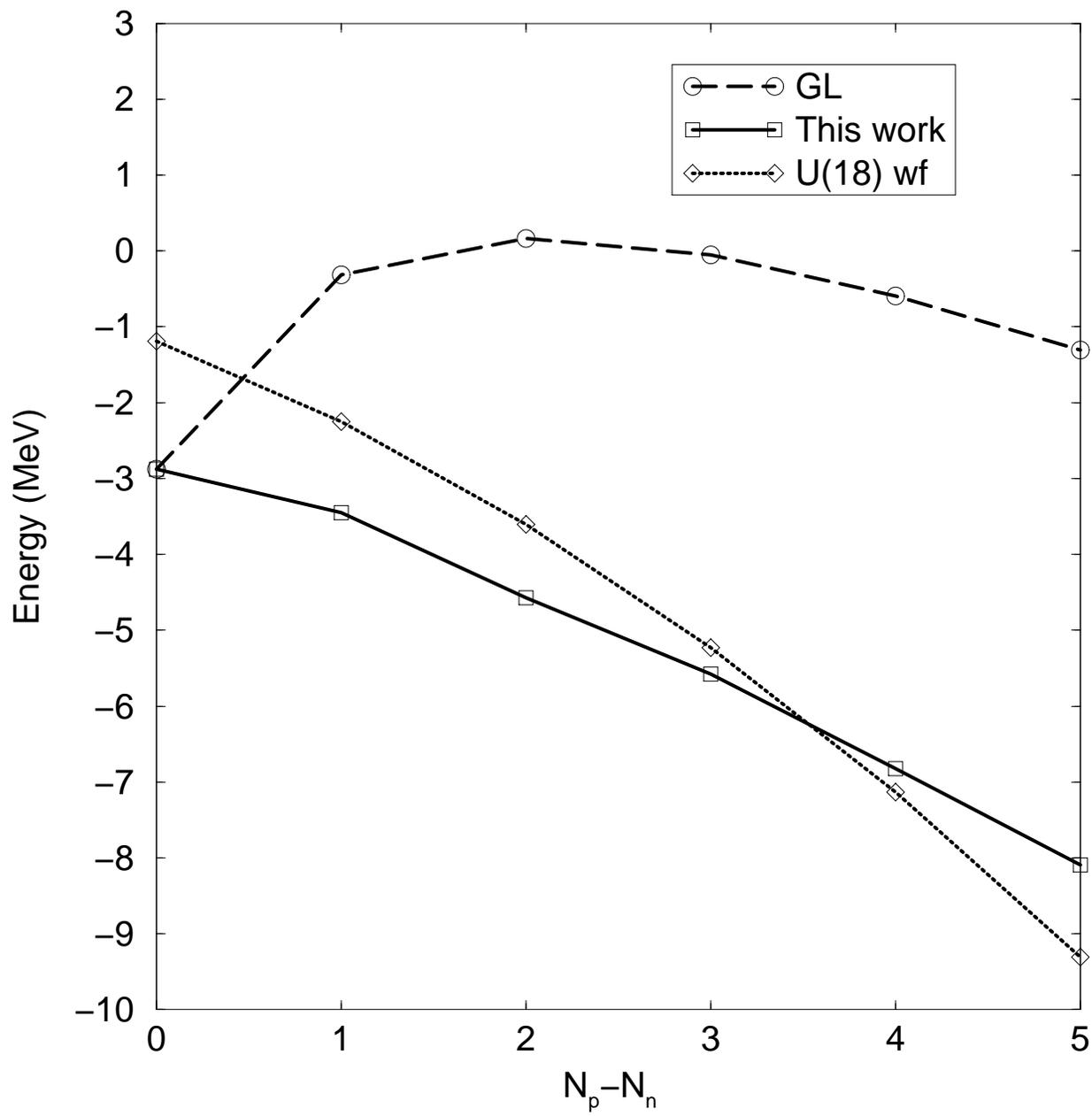
\end{document}